\documentclass[11pt,twoside]{article}

\usepackage{asp2014}

\aspSuppressVolSlug
\resetcounters

\begin{document}

\title{Using LSDB to enable large-scale catalog distribution, cross-matching, and analytics }

\author{Neven Caplar$^1,$$^3$, Wilson Beebe$^1$, Doug Branton$^1$, Sandro Campos$^2$, Andrew Connolly$^1$, Melissa DeLucchi$^2$, Derek Jones$^1$, Mario Juric$^1$, Jeremy Kubica$^2$, Konstantin Malanchev$^2$, Rachel Mandelbaum$^2$, Sean McGuire$^1$}
\affil{$^1$Department of Astronomy, University of Washington, Seattle, WA 98195, USA }
\affil{$^2$McWilliams Center for Cosmology and Astrophysics, Department of Physics, Carnegie Mellon University, Pittsburgh, PA 15213, USA}
\affil{$^3$ \email{ncaplar@uw.edu} }

\paperauthor{Neven~Caplar}{ncaplar@uw.edu}{0000-0003-3287-5250}{University of Washington}{Department of Astronomy}{Seattle}{WA}{98195}{USA}
\paperauthor{Wilson~Beebe}{wbeebe@uw.edu}{0009-0003-1791-8707}{University of Washington}{Department of Astronomy}{Seattle}{WA}{98195}{USA}
\paperauthor{Doug~Branton}{brantd@uw.edu}{0009-0009-7822-7110}{University of Washington}{Department of Astronomy}{Seattle}{WA}{98195}{USA}
\paperauthor{Sandro~Campos}{Author3Email@email.edu}{/0009-0007-9870-9032}{Carnegie Mellon University}{Department of Physics}{Pittsburgh}{PA}{15213}{USA}
\paperauthor{Andrew~Connolly}{ajc@astro.washington.edu}{0000-0001-5576-8189}{University of Washington}{Department of Astronomy}{Seattle}{WA}{98195}{USA}
\paperauthor{Melissa~DeLucchi}{delucchi@andrew.cmu.edu}{0000-0002-1074-2900}{Carnegie Mellon University}{Department of Physics}{Pittsburgh}{PA}{15213}{USA}
\paperauthor{Derek~Jones}{dtj1s@uw.edu}{0009-0006-2411-723X}{University of Washington}{Department of Astronomy}{Seattle}{WA}{98195}{USA}
\paperauthor{Mario~Juric}{mjuric@uw.edu}{0000-0003-1996-9252}{University of Washington}{Department of Astronomy}{Seattle}{WA}{98195}{USA}
\paperauthor{Jeremy~Kubica}{jkubica@andrew.cmu.edu}{0009-0009-2281-7031}{Carnegie Mellon University}{Department of Physics}{Pittsburgh}{PA}{15213}{USA}
\paperauthor{Konstantin~Malanchev}{kmalanch@andrew.cmu.edu}{0000-0001-7179-7406}{Carnegie Mellon University}{Department of Physics}{Pittsburgh}{PA}{15213}{USA}
\paperauthor{Rachel~Mandelbaum}{rmandelb@andrew.cmu.edu}{0000-0003-2271-152}{Carnegie Mellon University}{Department of Physics}{Pittsburgh}{PA}{15213}{USA}
\paperauthor{Sean~McGuire}{seanmcgu@andrew.cmu.edu}{0009-0005-8764-2608}{Carnegie Mellon University}{Department of Physics}{Pittsburgh}{PA}{15213}{USA}


\begin{abstract}
The Vera C.\ Rubin Observatory will generate unprecedented amounts of data, including $\sim$60 PB of raw data and $\sim$30 trillion observed sources, presenting a significant challenge for large-scale and end-user scientific analysis. As part of the LINCC Frameworks Project we are addressing these challenges with the development of the HATS (Hierarchical Adaptive Tiling Scheme) format and analysis package LSDB. HATS partitions data adaptively using a hierarchical tiling system to balance the file sizes, enabling efficient parallel analysis. Recent updates include improved metadata consistency, support for incremental updates, and enhanced compatibility with evolving datasets. LSDB complements HATS by providing a scalable, user-friendly interface for large catalog analysis, integrating spatial queries, crossmatching, and time-series tools while utilizing Dask for parallelization. We have successfully demonstrated the use of these tools with datasets such as ZTF and Pan-STARRS data releases on both cluster and cloud environments. We are deeply involved in several ongoing collaborations to ensure alignment with community needs, with future plans for IVOA standardization and support for upcoming Rubin, Euclid and Roman data. We provide our code and materials at \href{https://lsdb.io}{lsdb.io}.
\end{abstract}



\section{Motivation}

Vera C.\ Rubin Observatory, which has recently started commissioning, will produce data in amounts much larger than anything previously seen in optical astronomy \\ \citep{dmtn-135}. This includes around 60 PB of raw data, 40 billion observed stars, galaxies, and asteroids, and around 30 trillion observed sources. Distributing the data and large-scale practical scientific analysis with such datasets will be a challenge. \par 
The LINCC (LSST Interdisciplinary Network for Collaboration And Computing) Frameworks Project is an initiative by the LSST Discovery Alliance to enable such analysis. As part of this effort, we have been developing LSDB and HATS (Hierarchical Adaptive Tiling Scheme) format to deliver and enable end-user analysis on 10TB+ catalog datasets. 

\section{HATS  (Hierarchical Adaptive Tiling Scheme) format}

\begin{figure*}
    \centering
    \includegraphics[width=.99\textwidth]{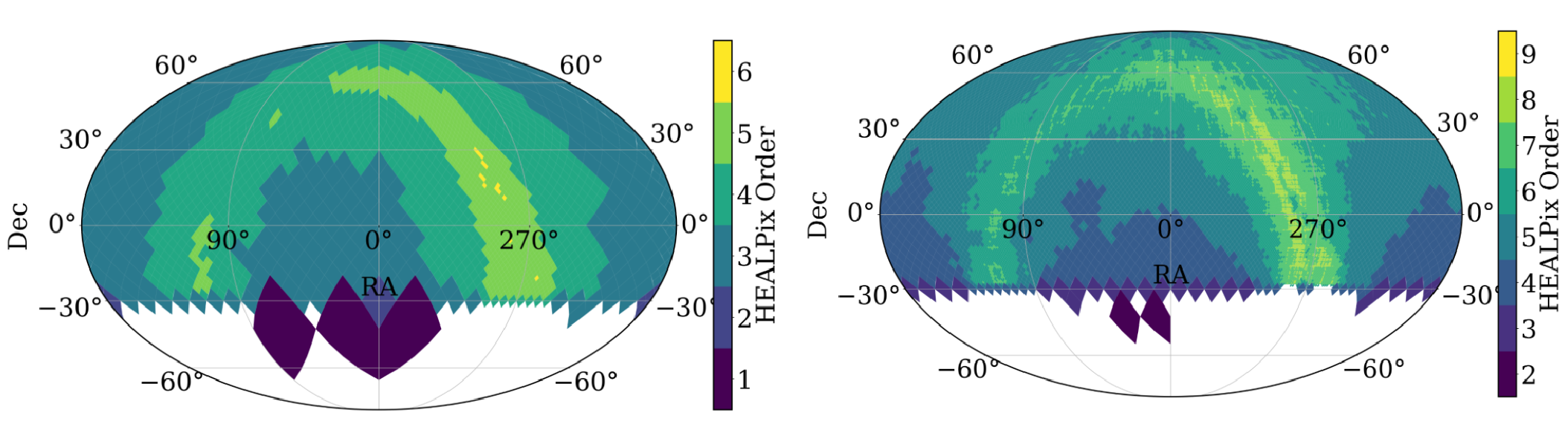}
    \caption{Partitioning of the ZTF data release 14 in the HATS scheme. \textit{Left panel}: Partitioning of the object table. \textit{Right panel}: Partitioning of the source table. The legend shows the healpix order of each pixel. In this Figure, ``source'' corresponds to each detection, while ``object'' is determined after associating nearby sources that likely correspond to the same astrophysical object. As such, there are more sources than objects in the catalog, necessitating deeper (higher value of healpix order) partitioning to get a similar size of all final pixels in memory. } 
    \label{fig:obj_src}
\end{figure*}

HATS provides a directory structure and associated metadata for spatially arranging large catalog survey data. We use healpix pixels at various orders to divide the sky into partitions, where each partition will have roughly the same number of objects. Healpix provides a natural way to split each pixel into four higher-order sub-regions that are equal in area. When constructing a catalog, we iteratively split the regions of the sky until each partition is just beneath some predefined threshold. In this way, all data partitions have comparable sizes, which allows for reasonable performance of parallel operations on each partition (see Figure \ref{fig:obj_src}).

We use Apache Parquet as the underlying storage format, as it provides efficient storage and retrieval of tabular data. Parquet files are highly structured, and each column is stored and compressed in a separate chunk. If users only require a few columns at a time for their analysis, we can read only those specific columns, reducing input and output operations as well as overall memory usage. Parquet has a robust ecosystem of libraries for reading, writing, and analyzing files, so catalog providers can be assured that their users will have their pick of suitable tools. \par

The basis of this format was presented at ADASS 2023 \citep{Wy23_adassxxxii}. We want to emphasize here the changes and improvements since that iteration of the code. First, we have renamed the package and format from HiPSCat to HATS. We have done this to avoid potential confusion with the HiPS format and HiPS Catalogs, an IVOA standard from which we borrow our directory structure. As part of that effort, we have also conducted an effort to harmonize our metadata files and field names to be more consistent with the original HiPS format properties file. Furthermore, we have replaced the spatially-generated unique index that was required in the previous iteration of the code with a general-purpose spatial index based on healpix position at order 29 (the highest order index that fits in a 64 bit integer). We did this to continue to provide fast spatial operations and global ordering  while relaxing uniqueness limitations.
Finally, we have also refined supplemental tables, including storage of secondary index lookups, pixel margin cache, and non-point-source region data like dust maps and pixel masks.

\section{LSDB}

\begin{figure*}
    \centering
    \includegraphics[width=.49\textwidth]{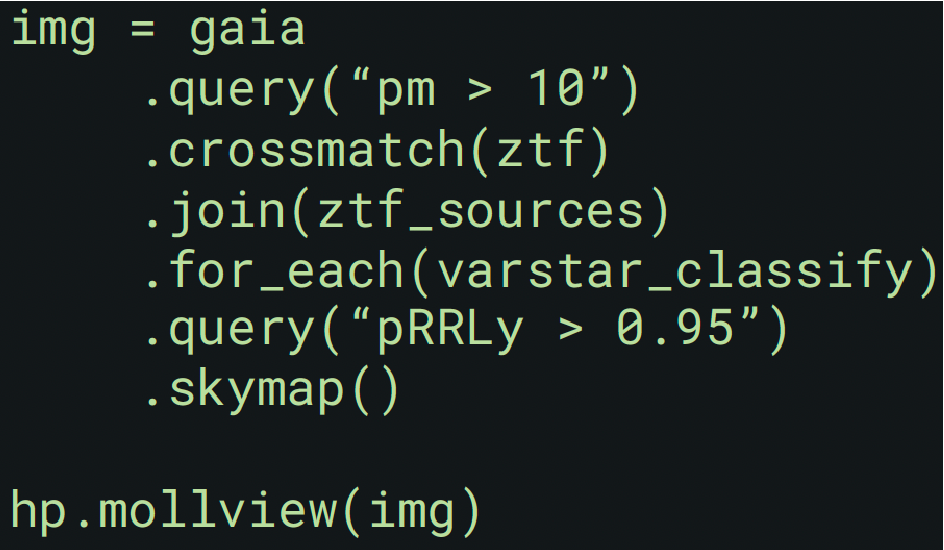}
    \caption{Targeted API for the LSDB code. This example shows a hypothetical scientific analysis of selecting starts from the GAIA survey with high proper motion, cross-matched to ZTF data release. We then extract lightcurves, classify them and choose the likely RR Lyrae variable stars, before representing their distribution on the sky. LSDB aims to make such whole-sky analysis achievable and easy for science end-users.  } 
    \label{fig:api}
\end{figure*}

In conjunction with the format development, we have been developing a user-oriented package to enable scientific use of catalogs stored in HATS format. Our main goal is to enable large, all-sky analysis workflows, as illustrated in the code block shown in Figure \ref{fig:api}. LSDB enables Pandas-like analysis with a large number of cores and uses Dask to support parallelization while keeping full HATS awareness, such as enabling spatial queries, cross-matching, time-series analysis, and multi-dataset joining.\par

In particular, since last year's proceedings, we have spent considerable time enabling large time-domain analysis. The challenge with the original implementation lay in the massive memory usage and computational overhead caused by the continuous synchronization and joining of metadata and photometry catalogs representing time-domain data. To address this, we developed a new library, \texttt{nested-pandas}\footnote{\href{https://github.com/lincc-frameworks/nested-pandas}{https://github.com/lincc-frameworks/nested-pandas}}, which pre-joins light-curve data into a compact representation: a single pandas column. Each element of this column resembles a nested data frame, representing a single light curve, while the entire column is efficiently stored as an Arrow structured array. With this implementation, and its extension (\texttt{nested-dask}\footnote{\href{https://github.com/lincc-frameworks/nested-dask}{https://github.com/lincc-frameworks/nested-dask}}) we successfully ran extensive analysis pipelines. For instance, we extracted a low-resolution periodogram for a billion light curves from Zwicky Transient Factory (ZTF) Data Release 14 \citep{2019PASP..131a8002B} in two hours using seven nodes of the Bridges2 Supercomputer Cluster. \par

\section{Collaboration and community building}
We have also started providing publicly available HATS catalogs online, primarily via the webpage \href{https://data.lsdb.io}{data.lsdb.io}. This data is hosted on servers operated by the University of Washington. We have also worked with STScI (Space Telescope Institute) and IPAC (Infrared Processing \& Analysis Center) to provide their large catalogs (primarily PanSTARRS \citep{2020ApJS..251....7F} and ZTF data releases) publicly online in HATS format. We have conducted experiments with the earlier versions of the HATS format, and we expect to provide consistent access to these catalogs via their cloud resources. Similarly, our partners at Strasbourg Astronomical Data Center (CDS) have created an experimental online access point for their implementation of the HATS format with the existing catalogs they provide. \par
An essential aspect of our effort is to ensure that our existing code is working seamlessly on cloud platforms, given the considerable interest of the community and the goals expressed in the Decadal survey \citep{astro2020}. We have tested our workflows on the Fornax platform, cloud environment being developed by NASA.

As an example of successful collaboration, we want to emphasize the successful implementation of our scheme with the SPlus survey \citep{2019MNRAS.489..241M}, which implemented the HATS format as their primary avenue to provide bulk downloads. Their engineers also provided an innovative way to filter data, as queries can be evaluated on the server to reduce the download size needed from the users, and we plan to implement this functionality more widely. \par
Finally, we are in contact with the GAIA consortium, who plan to provide GAIA data release 4 bulk download in parquet format, using a modification of the HATS format; and we aim to provide a translation layer between the two formats for ease of use.

\section{Conclusions and Future plans}

Rubin Observatory recently started taking its first images and is expected to ramp up scientific imaging in the first half of 2025. We are ready to support the commissioning efforts and provide bulk access to the data as it becomes available. We will also continue our collaboration efforts and aim to provide more catalogs with our partners in HATS format. These efforts will also include preparation for data releases for Euclid and Roman missions. \par
As a part of this effort, we aim to apply for IVOA endorsed note/standarization. Given the positive feedback we have received and excellent communication with our partners, we are confident that we are going to be able to start the process in 2025.

\acknowledgements  This project is supported by Schmidt Sciences. This work used Bridges-2 at Pittsburgh Supercomputing Center, provided through ACCESS program.

\bibliography{F601}

\end{document}